\documentclass[%
 aip,
 amsmath,amssymb,
 reprint,%
]{revtex4-1}

\usepackage{graphicx}
\usepackage{dcolumn}
\usepackage{bm}

\usepackage[utf8]{inputenc}
\usepackage[T1]{fontenc}
\usepackage{mathptmx}
\usepackage{etoolbox}

\makeatletter
\def\@email#1#2{%
 \endgroup
 \patchcmd{\titleblock@produce}
  {\frontmatter@RRAPformat}
  {\frontmatter@RRAPformat{\produce@RRAP{*#1\href{mailto:#2}{#2}}}\frontmatter@RRAPformat}
  {}{}
}%
\makeatother
\begin{document}

\preprint{AIP/123-QED}

\title{Improved performance of polycrystalline antiferromagnet/ferromagnet stack by nitrogen assisted deposition}

\author{Y. Khaydukov}
\email{Yury.Khaydukov@Seagate.com}
\affiliation{ 
Seagate Technology, Derry, Northern Ireland, BT48 0BF, UK
}%

\author{G. McCafferty}%
\affiliation{ 
Seagate Technology, Derry, Northern Ireland, BT48 0BF, UK
}%

\author{A. Dobrynin}%
\affiliation{ 
Seagate Technology, Derry, Northern Ireland, BT48 0BF, UK
}%

\author{A. Devishvili}%
\affiliation{ 
Institute Laue Langevin, Grenoble, France
}%

\author{A.Vorobiev}%
\affiliation{ 
Department of Physics and Astronomy, Uppsala University, Uppsala, Sweden
}%

\author{P.Bencok}%
\affiliation{ 
Diamond Light Source, Didcot, Oxfordshire, OX11 0DE, UK
}%

\author{R.Fan}%
\affiliation{ 
Diamond Light Source, Didcot, Oxfordshire, OX11 0DE, UK
}%

\author{P.Steadman}%
\affiliation{ 
Diamond Light Source, Didcot, Oxfordshire, OX11 0DE, UK
}%

\author{K. McNeill}%
\affiliation{ 
Seagate Technology, Derry, Northern Ireland, BT48 0BF, UK
}%

\author{M. Ormston}%
\affiliation{ 
Seagate Technology, Derry, Northern Ireland, BT48 0BF, UK
}%

\date{\today}

\begin{abstract}
The deposition process of the IrMn$_3$/Co$_{70}$Fe$_{30}$ bilayer of antiferromagnet/ferromagnet (AF/FM) type was modified by introducing a nitrogen additive in argon plasma during the magnetron sputtering of the Co$_{70}$Fe$_{30}$ layer. This slight modification significantly enhanced the exchange bias energy of the AF/FM bilayer for an IrMn layer thickness ranging from 20{\AA} to 50{\AA} and reduced the coercivity of the FM layer. Calculations indicate that the boost in exchange bias energy and the reduction in coercivity can be attributed to the increased anisotropy energy of the antiferromagnet, resulting in more effective pinning of the ferromagnet by antiferromagnetic grains. The increase in anisotropy is caused by the diffusion of nitrogen from the FM into the AF layer, as established by X-ray diffraction, Neutron Reflectometry, and X-ray Magnetic Circular Dichroism. Our research allows to improve magnetic characteristics of exchange-coupled FM/AF structures through minor modifications in the sputtering process and/or save up to 20\% of the costly IrMn$_3$ target by reducing the thickness of the AF layer.

\end{abstract}

\maketitle

The exchange bias effect between antiferromagnetic (AF) and ferromagnetic (FM) layers is widely used in various spintronic applications to pin the direction of the FM layer’s magnetic moment. In typical industrial applications, a polycrystalline disordered  $\gamma$-IrMn$_3$ is used as an AF layer. The polycrystallinity of the AF layer means it is composed of small crystallites (grains) of different sizes. Some small grains, due to their low magnetocrystalline anisotropy, are not pinned and can rotate with the FM layer. This results in reduced interfacial exchange coupling (IEC) and increased coercivity, as the anisotropies of the reversible AF grains contribute to the FM’s energy barrier. Furthermore, the presence of such small grains with low anisotropy increases AF thermal glitching, resulting in low-frequency magnetic noise that affects the device’s signal-to-noise ratio. Another issue is the thermal stability of the AF/FM interface during the annealing process: to set the magnetization of FM layer in right direction one needs to anneal the stack at a temperature above the blocking temperature $T_b$ of the AF layer in an applied magnetic field. If the annealing temperature is too high, interdiffusion processes at the AF/FM interface can lead to degradation of AF/FM bilayer performance \cite{Lee02,Lee02(2),Quarterman19,Khaydukov24}. To improve thermal stability upon annealing it was proposed to use very thin ($\sim$ 1\AA) Ru layers between IrMn$_3$ and Co$_{70}$Fe$_{30}$ layer \cite{Nikolaev16}. Experiment has shown that this thin layer allows to improve thermal stability of the bilayer, however at the cost of 2-3 times reduced IEC energy density.

Series of works reported modification of properties of AF or FM layers by using reactive magnetron sputtering in partial atmosphere of nitrogen. In particular, it was demonstrated that the nitrogen-assisted deposition (NAD) of CoFe alloy leads to amorphization and magnetic softening of the layers \cite{Han23,Hwang16,Kuo98,Sun00,Wu15,Xu13,Zhang_2012}. In a series of works of Meinert et al. \cite{Meinert15,Zilske17,Dunz17,Callori16}  it was demonstrated that deposition of Mn in nitrogen atmosphere leads to appearance of new antiferromagnetic MnN layer with high Neel temperature of $T_N$=660 K and high coupling strength to the adjacent CoFe ferromagnetic layer ($J_{eb}$=0.41 mJ/m$^2$). 

\begin{figure*}
\includegraphics[width=1.0\textwidth]{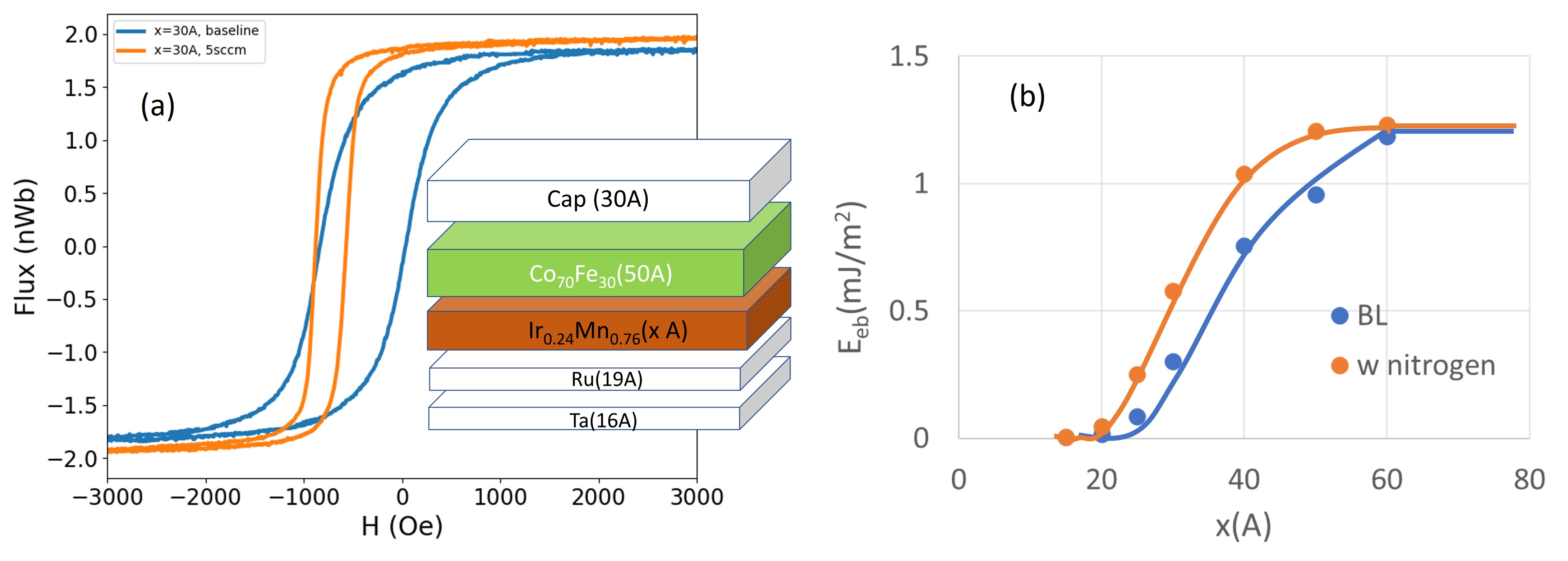} 
\caption{\label{fig:Magn} (a) Hysteresis loops measured on the sample with x=30\AA. Inset shows a sketch of the structures used for the study. (b) Exchange bias energy density versus thickness of the AF layer for the baseline (blue dots) and samples with nitrogen assisted deposition (orange dots). Solid lines are drawn to guide the eye.}
\end{figure*}

In this work we set out to explore as nitrogen assisted deposition impacts structural and magnetic properties of IrMn$_3$/Co$_{70}$Fe$_{30}$ stack of AF/FM type widely used in industrial applications. For the experiment we have deposited a series of Ta(16\AA)/Ru(19\AA)/IrMn$_3$(x)/Co$_{70}$Fe$_{30}$(50\AA) structure capped with a 30{\AA} layer Ru layer (inset in Fig.~\ref{fig:Magn}a) using magnetron deposition technique. The antiferromagnetic IrMn$_3$ (IM) layer was deposited at 275$^\circ$C on an optimally chosen Ta/Ru seed. Next, the FM Co$_{70}$Fe$_{30}$ (CF) layer was deposited at room temperature. Reference samples were deposited in argon atmosphere with a typical flow of 40 sccm (standard cubic centimeters per minute). Samples under the study were deposited using additional flow of nitrogen in the CF layers. We will further denote samples as CFyN, where $y$ indicates the flow of nitrogen in sccm in addition to the standard Ar flow. 

Blue line in Fig.~\ref{fig:Magn}a shows hysteresis loop for the reference sample with x=30\AA. The loop is characterized by coercive field $H_c$ = 430 Oe and exchange bias field $H_{eb}$= -400 Oe, which corresponds to the IEC energy density of $E_{eb}$ = 0.3 mJ/m$^2$. Using of additional 5 sccm of nitrogen during deposition of CF layer leads to 40\% decrease of coercive field to $H_c$ = 250 Oe and 80\% increase of exchange bias field to $H_{eb}$= - 725 Oe ($E_{eb}$ = 0.5 mJ/m$^2$). The same procedure was performed for the other samples with different $x$. Fig.~\ref{fig:Magn}b summarizes the $E_{eb}$(x) dependence for reference and nitrogenated samples. As one can see, for intermediate thicknesses 20{\AA} <$x$ <60{\AA} one can see a 0.2-0.3 mJ/m$^2$ increased exchange energy of nitrogenated samples with respect to the reference sample.

\begin{figure}
\includegraphics[width=0.5\textwidth]{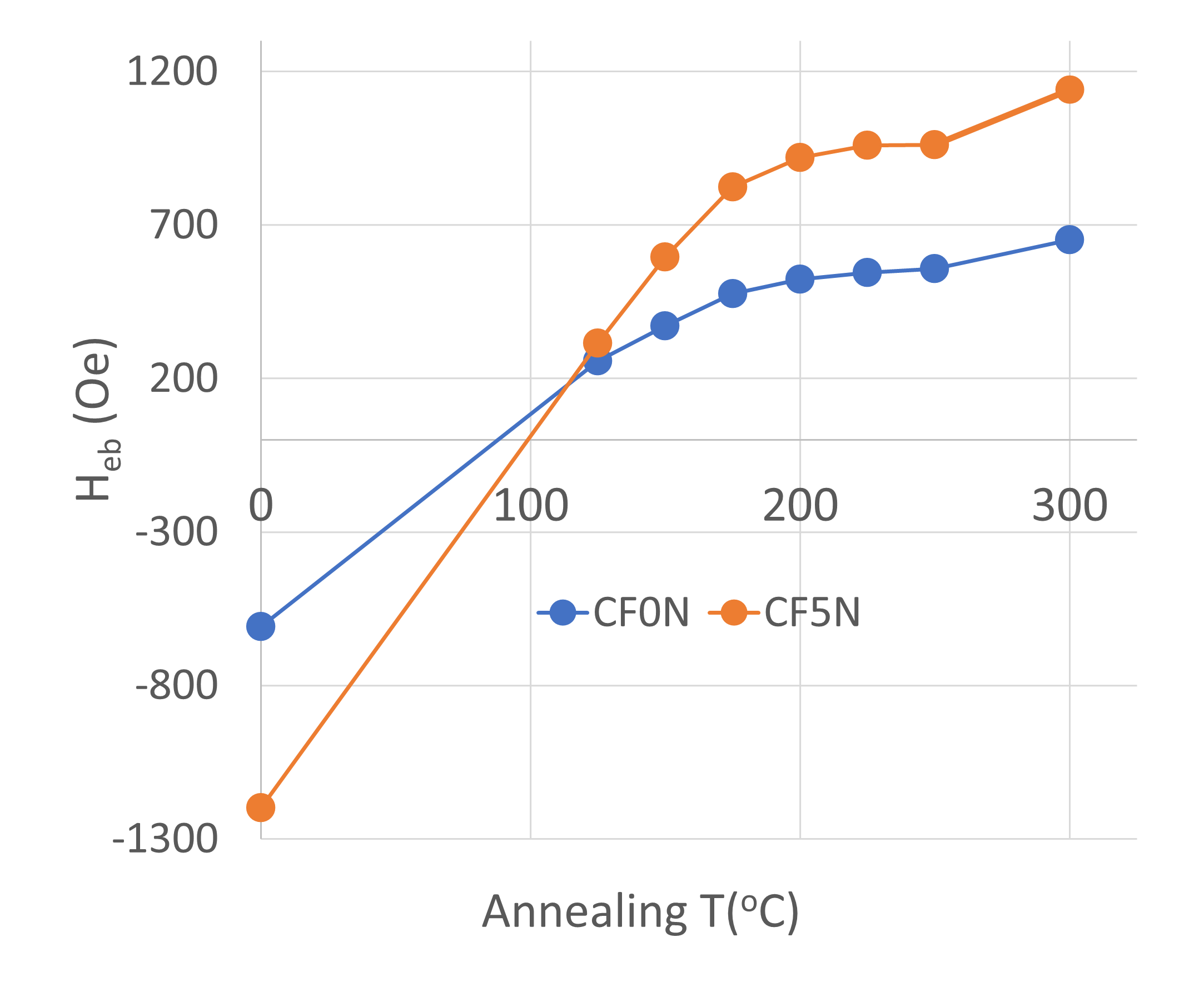} 
\caption{\label{fig:Tdep} Dependence of the exchange bias field on the annealing temperature for the structure IM(30\AA)/CF(50\AA) deposited under standard conditions (blue) and with addition of 5 sccm of nitrogen (orange).}
\end{figure}

Fig.~\ref{fig:Tdep} shows the dependence of the exchange bias on the annealing temperature for the structure with $x$=30{\AA} measured using protocol described in Ref.\cite{Soeya94}. It demonstrates that the exchange bias field of the nitrogenated sample is (a) systematically higher than the reference sample and (b) keeps monotonously increasing with temperature while the baseline stopped growing above 250$^\circ$C.

\begin{figure}
\includegraphics[width=0.5\textwidth]{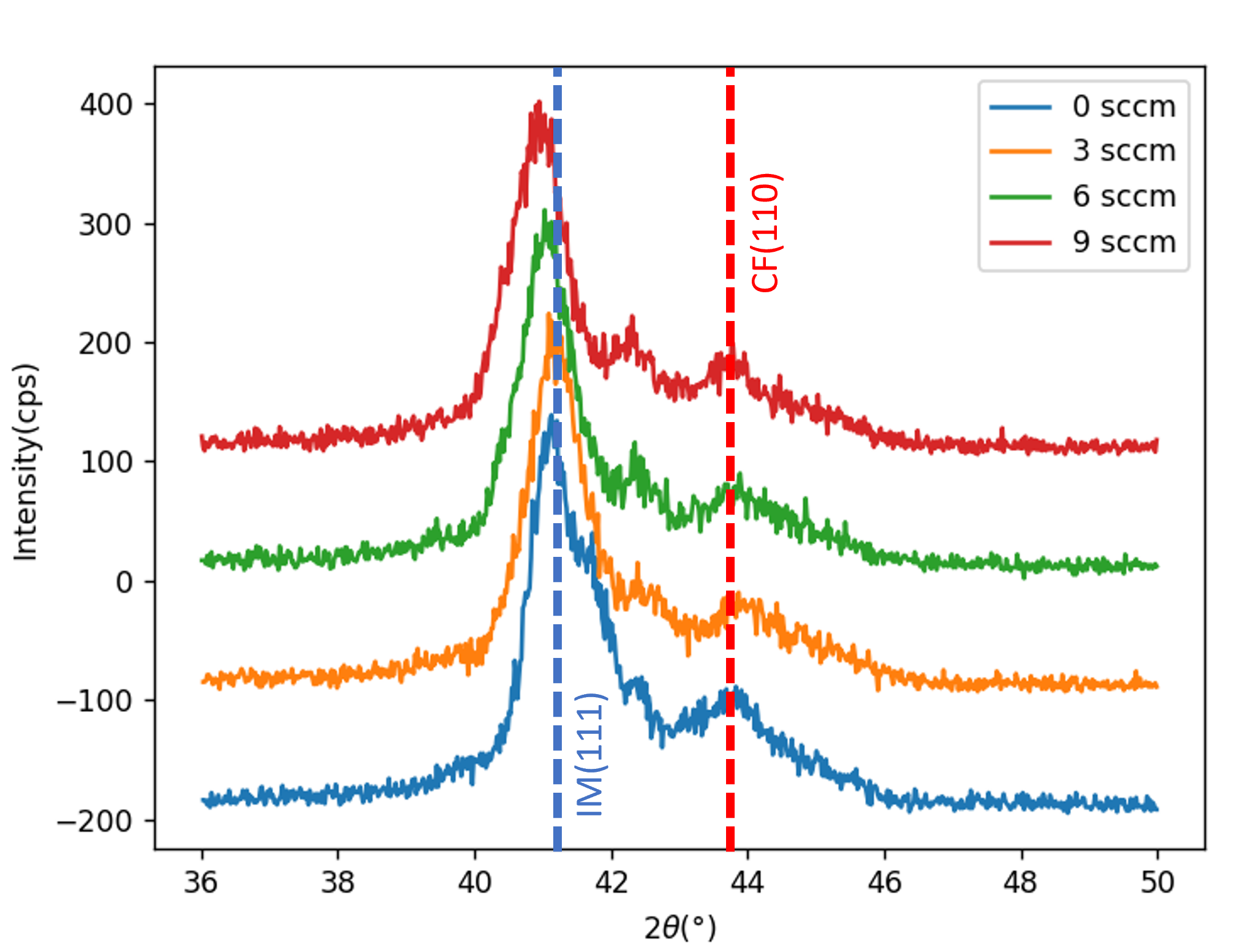} 
\caption{\label{fig:XRD} The X-ray diffraction pattern for family of samples with x=60\AA.}
\end{figure}

To shed light on the structural changes leading to the observed magnetic effects, we conducted a series of X-ray and neutron experiments. Fig.~\ref{fig:XRD} shows $\theta$/2$\theta$ X-ray diffraction (XRD) pattern of the family of samples with x=60\AA. For the reference sample (blue line) two distinct peaks can be seen at 2$\theta$=41.3$^\circ$ and 2$\theta$=43.8$^\circ$ corresponding to IM(111) and CF(110) peaks respectively. For the nitrogenated samples we have observed a linear shift of the IM(111) peak towards smaller angles which would mean an expansion of the IM lattice. At the same time, we have not observed any systematic change of the position of the CF peak. Such a change in diffraction patterns may mean the migration of nitrogen from the CF into the IM crystal lattice, leading to its expansion.

\begin{figure*}[t]
\includegraphics[width=1.0\textwidth]{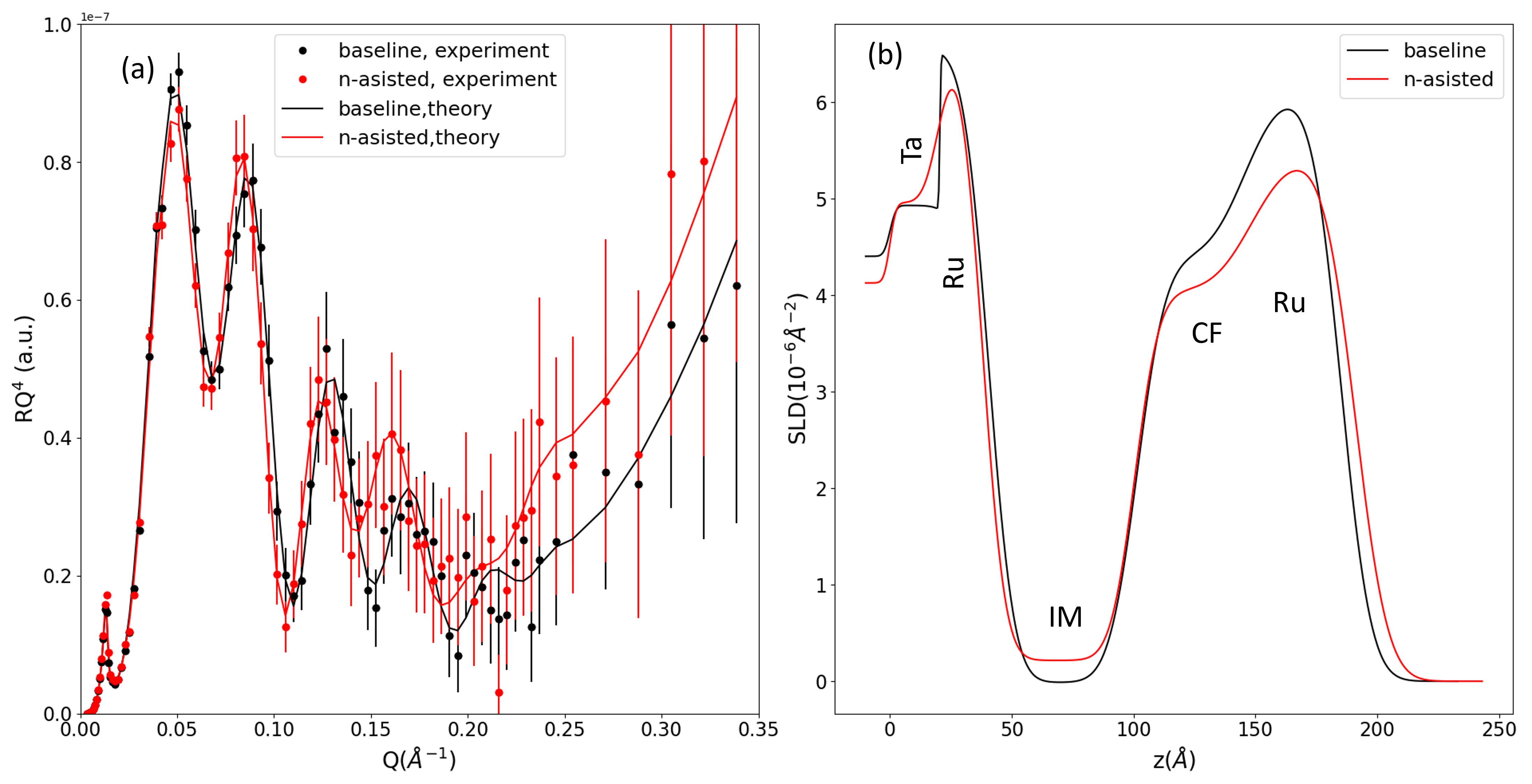} 
\caption{\label{fig:NR} (a) Experimental (dots) and model (solid lines) Q$^4$-normalized reflectivities  of the reference CF0N (black) and nitrogenated CF3N samples with x=60\AA. (b) Corresponding nuclear SLD depth profiles of the reference (black) and nitrogenated (red) samples.}
\end{figure*}

Additional confirmation of nitrogen migration into the AF layer was obtained by neutron reflectometry measured at the reflectometer Super ADAM \cite{Devishvili2013,Vorobiev2015}. Figure ~\ref{fig:NR}a shows two non-polarized reflectivities for samples with x=60{\AA} (reference and with 3 sccm of N$_2$). The nuclear scattering length density (SLD) profiles obtained as a result of the fit in the genX program are shown in Fig. ~\ref{fig:NR}b. It can be seen that nitrogenation of the sputtering process leads to a fairly strong change in all layers. So, the SLD of the AF layer increased from -0.02$\times10^{-6}$ \AA$^{-2}$ to 0.4$\times10^{-6}$ \AA$^{-2}$ with an increase in thickness from 6.1 nm to 6.5 nm. Change of the SLD and increase of thickness is similar to the observed earlier hydrogen incorporation into thin Nb or V films \cite{Callori16,Guasco22,Perrichon22}. A pronounced contrast between the average scattering length of IM and nitrogen allows for the estimation of the nitrogen concentration in the IM layer as $\sim$10\%, in assumption of constant layer density. Changes in other layers can be caused by both the migration of nitrogen and the secondary diffusion of other atoms provoked by nitrogen. However, given the weaker contrast, quantitative analysis should involve complementary techniques which is out of the scope of this work. 

\begin{figure*}
\includegraphics[width=1.0\textwidth]{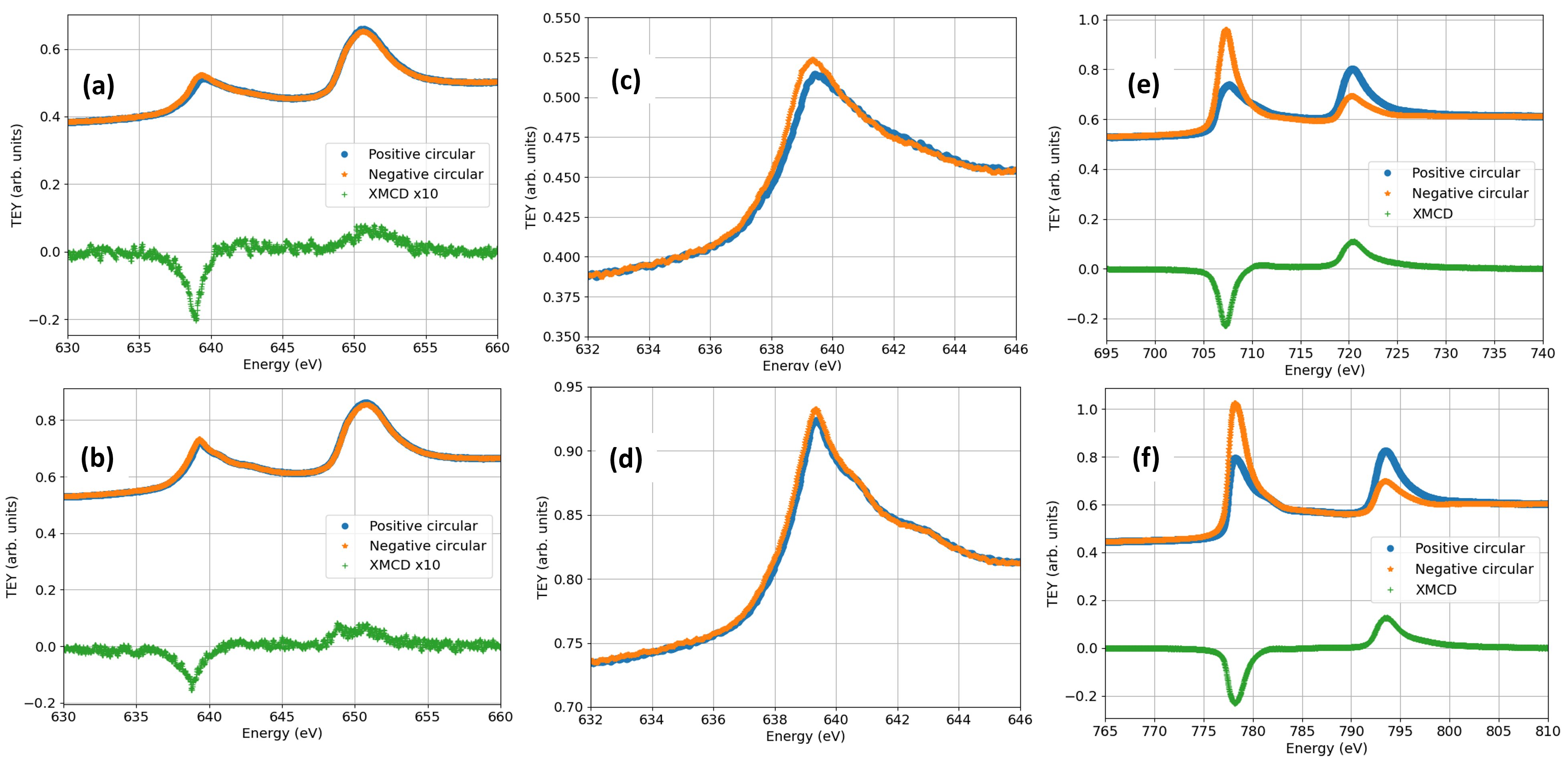} 
\caption{\label{fig:XMCD} 
XAS and XMCD spectra measured across Mn L2/3 absorption edges on the reference sample CF0N (a) and nitrogenated sample CF3N (b); zoomed in XAS spectra at the Mn L3 absorption edge for samples CF0N (c) and CF3N (d); XAS and XMCD for sample CF3N measured across Fe L2/3 (e) and Co L2/3 (f) absorption edges.
}
\end{figure*}

Soft X-ray magnetic circular dichroism (XMCD) experiments were performed at beamline I10, Diamond Light Source. X-ray absorption spectra (XAS) at Mn, Fe, and Co L2/3 absorption edges were measured using total electron yield detection in grazing (in-plane) and normal (out of plane) geometries. Fig. ~\ref{fig:XMCD}  shows some spectra taken from reference (a), (c) and nitrogenated (b),(d),(e),(f) samples at the grazing incidence with saturation (1T) field applied in-plane, (nearly) collinear with the incident beam direction. It can be seen that there is a significant uncompensated moment at the Mn edges for both samples, which is aligned with the CF magnetizations direction, as the XMCD signals at Mn, Fe, and Co edges have the same sign. Most pronounced difference between the two samples is in the emergent fine structure at the L3 Mn edge for the nitrogenated samples (Fig.~\ref{fig:XMCD}d). This multiplet structure implies changed chemical environment for the Mn. While quantitative analysis of this structure would require more detailed XANES measurements and modelling, qualitatively the changed Mn environment is an indication of the presence of N impurities, affecting the Mn symmetry, and therefore properties of the IM layer. In case of the normal incidence a higher field (2T) was applied to saturate the sample in the out-of-plane geometry. While the XAS and XMCD spectra at the Fe and Co edges look identical to the grazing incidence (not shown), for the Mn edges there was no XMCD signal observed. This implies that the observed uncompensated Mn moment is confined in plane. While this observation is not directly relevant to the topic of this work, we find it important to highlight, as this effect deserves further investigation.

All experimental data described above indicate a change in the magnetic behavior of the IM/CF structures caused by the migration of nitrogen from the CF to the IM layer. The XRD data indicate that nitrogen is incorporated into the IM interstitial states expanding its lattice. It seems reasonable to suggest that this lattice distortion leads to change of magnetocrystalline anisotropy of the AF layer. The effect of the increased AF anisotropy on the exchange bias and coercivity in a polycrystalline AF/FM bilayer can be assessed using the following model.  It’s known that finite size effects lead to AF magnetocrystalline anisotropy constant $K_u$ dependence on the grain size \cite{Mishra10}. Reformulating the experimental thickness dependence of the AF anisotropy \cite{Mishra10} to area dependence at a fixed AF film thickness, we can write: $K_u(A) = \frac{K_u(\infty)}{1 - \left(\frac{A_0}{A}\right)^\gamma}$, where $\gamma$=1.07, $A_0$=10 nm$^2$. Here the scaling exponent $\gamma$ is a 2D equivalent of the one found in Ref.\cite{Mishra10}  for the 1D case (thickness of the IM layer), and parameter $A_0$  defines the grain size at which anisotropy fully disappears. It has been found that $K_u=5.5\times10^5  J/m^3$  for 4 nm thick IrMn$_3$ film \cite{Vallejo07}.

From simple FM layer’s (approximated with a macrospin here) energy barrier considerations, it’s easy to show that left and right coercive fields of the FM/AF bilayer  $H_{cL}$ and $H_{cR}$ can be found from the following expressions:

\begin{equation}
\label{eq:eq1}
-\mu_0 M_s A_{\text{tot}} t_{\text{FM}} H_{cL} = t_{\text{AF}} \sum_{\text{mob}} A_i K_{u_i} + \sigma \sum_{\text{fix}} A_j
\end{equation}

\begin{equation}
\label{eq:eq2}
\mu_0 M_s A_{\text{tot}} t_{\text{FM}} H_{cR} = t_{\text{AM}} \sum_{\text{mob}} A_i K_{u_i} - \sigma \sum_{\text{fix}} A_j
\end{equation}
Here $\mu_0 M_s$=2T is the FM layer’s magnetization, $t_{FM}$=10nm - thickness of the FM layer, $t_{AF}$=3nm - thickness of the AF layer  $\sigma$=1.2 mJ/m$^3$ – IEC energy density, $\sum_{\text{mob}} A_i$- sum of the product of the grain’s surface area by its anisotropy constant over reversible AF grains, $\sum_{\text{fix}} A_i$ - sum of the fixed AF grains’ surface areas. The condition for the reversibility of the i$^{th}$ AF grain is $K_{u_i} t_{\text{AF}} < \sigma$.

\begin{figure*}
\includegraphics[width=1.0\textwidth]{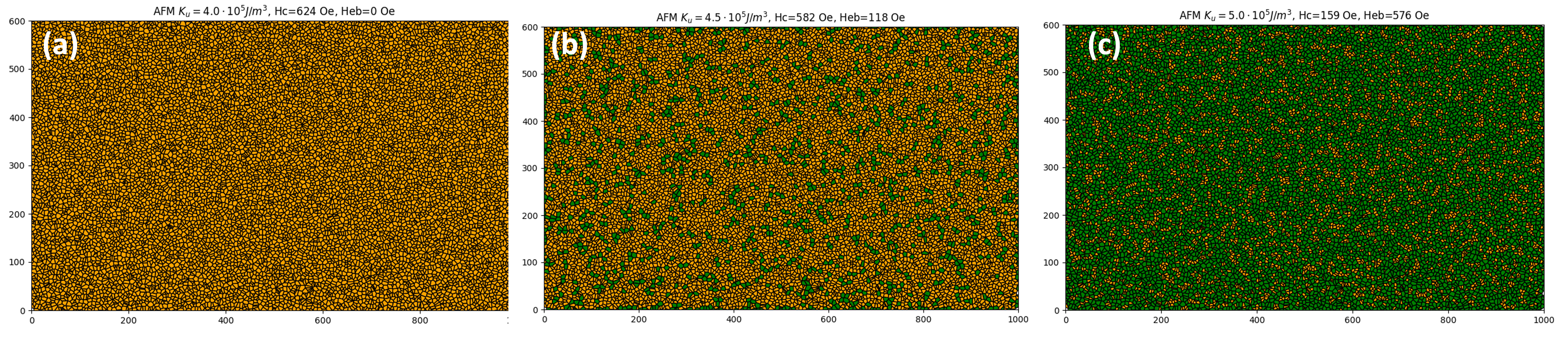} 
\caption{\label{fig:Voron} 
Top view of a polycrystalline AF layer, approximated by Voronoi cells. Orange cells – reversible AF grain, contributing to coercivity, green cells – fixed AF grains, contributing to exchange bias (a) $K_u=4.0\times10^5  J/m^3$ (b) $K_u=4.5\times10^5  J/m^3$ (c) $K_u=5.0\times10^5  J/m^3$
}
\end{figure*}

Fig. ~\ref{fig:Voron}a-c show the Voronoi generated AF layer surface with the average grain size of 7 nm at $K_u=4.0,4.5,5.0\times10^5  J/m^3$  respectively. In these figures, the orange-colored grains are mobile, while the green ones are fixed. One can see that an increase in the anisotropy constant $K_u$ leads to a significant suppression of the concentration of reversible grains. 

From Eq.\ref{eq:eq1} and Eq.\ref{eq:eq2} we obtain the expressions for coercivity and exchange bias fields:

\begin{equation}
\label{eq:eq3}
H_c = \frac{t_{\text{AF}} \sum_{\text{mob}} A_i K_{u_i}}{\mu_0 M_s A_{\text{tot}} t_{\text{FM}}}
\end{equation}

\begin{equation}
\label{eq:eq4}
H_{EB} = -\frac{\sigma \sum_{\text{fix}} A_j}{\mu_0 M_s A_{\text{tot}} t_{\text{FM}}}
\end{equation}

\begin{figure}
\includegraphics[width=0.5\textwidth]{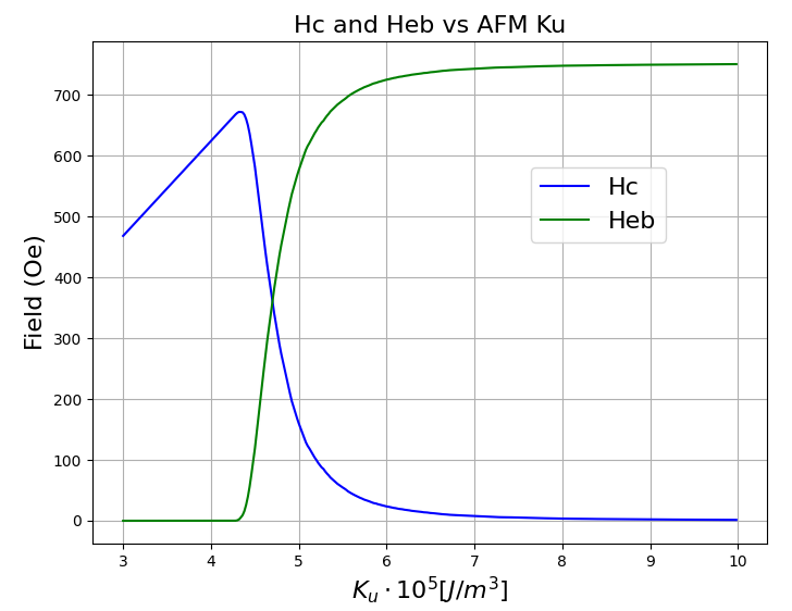} 
\caption{\label{fig:Rslt} 
Coercivity (blue line) and exchange bias field (green line) as a function of AF anisotropy constant.
}
\end{figure}

Resulting plot of these fields vs the AF magnetocrystalline anisotropy constant for the above parameters is shown in Fig. ~\ref{fig:Rslt}. From these results it is clear that the AF anisotropy increase leads to the decrease of coercivity and increase of exchange bias, which is consistent with experimental observations.

The obtained process correction with addition of nitrogen can be applied as follows to improve the AF/FM stack performance:
\begin{itemize}
    \item One can keep the same thickness as in the standard process. In this case the NAD of CF layer will lead to a stronger pinning by the higher anisotropy IM layer. This, in turn, will decrease magnetic noise from the pinned layer through its effective stiffness increase. As an example, we show in Fig. ~\ref{fig:Magn}a two hysteresis loops for $x$=30\AA, where deposition at additional 8 sccm nitrogen leads to increase magnitude of exchange bias field from 0.4 kOe to 0.7 kOe and suppression of coercivity from 0.4 kOe to 0.2 kOe.
    
    \item Alternatively, one can keep the same pinning strength and thermal stability level (hence same AF glitching) as in a standard process, but have thinner IM layers with higher anisotropy, prepared by the NAD. That will allow to (a) reduce device dimensions and (b) decrease consumption of expensive IM targets. From Fig. ~\ref{fig:Magn}b one can see that for example exchange energy of nitrogenated system with x=50{\AA} is the same as $x$=60{\AA}  of standard process. So the stack with new process will have 10{\AA} smaller thickness and consumption of the IM target would be 20\% less.

\end{itemize}

In summary, we modified the IrMn($x$)/CoFe bilayer deposition process by introducing a small nitrogen content during CoFe layer growth. This adjustment resulted in an increased interfacial exchange coupling energy density and a reduced coercive field for stacks with 20\AA <$x$< 60\AA. We attribute these changes to the enhanced magnetocrystalline anisotropy of small grains, facilitated by nitrogen migration from the CoFe to the IrMn layer. Our study holds promise for optimizing spintronic devices that exploit the exchange bias effect through subtle deposition process variations.

\bibliography{aipsamp}

\begin{thebibliography}{23}%
\makeatletter
\providecommand \@ifxundefined [1]{%
 \@ifx{#1\undefined}
}%
\providecommand \@ifnum [1]{%
 \ifnum #1\expandafter \@firstoftwo
 \else \expandafter \@secondoftwo
 \fi
}%
\providecommand \@ifx [1]{%
 \ifx #1\expandafter \@firstoftwo
 \else \expandafter \@secondoftwo
 \fi
}%
\providecommand \natexlab [1]{#1}%
\providecommand \enquote  [1]{``#1''}%
\providecommand \bibnamefont  [1]{#1}%
\providecommand \bibfnamefont [1]{#1}%
\providecommand \citenamefont [1]{#1}%
\providecommand \href@noop [0]{\@secondoftwo}%
\providecommand \href [0]{\begingroup \@sanitize@url \@href}%
\providecommand \@href[1]{\@@startlink{#1}\@@href}%
\providecommand \@@href[1]{\endgroup#1\@@endlink}%
\providecommand \@sanitize@url [0]{\catcode `\\12\catcode `\$12\catcode `\&12\catcode `\#12\catcode `\^12\catcode `\_12\catcode `\%12\relax}%
\providecommand \@@startlink[1]{}%
\providecommand \@@endlink[0]{}%
\providecommand \url  [0]{\begingroup\@sanitize@url \@url }%
\providecommand \@url [1]{\endgroup\@href {#1}{\urlprefix }}%
\providecommand \urlprefix  [0]{URL }%
\providecommand \Eprint [0]{\href }%
\providecommand \doibase [0]{http://dx.doi.org/}%
\providecommand \selectlanguage [0]{\@gobble}%
\providecommand \bibinfo  [0]{\@secondoftwo}%
\providecommand \bibfield  [0]{\@secondoftwo}%
\providecommand \translation [1]{[#1]}%
\providecommand \BibitemOpen [0]{}%
\providecommand \bibitemStop [0]{}%
\providecommand \bibitemNoStop [0]{.\EOS\space}%
\providecommand \EOS [0]{\spacefactor3000\relax}%
\providecommand \BibitemShut  [1]{\csname bibitem#1\endcsname}%
\let\auto@bib@innerbib\@empty
\bibitem [{\citenamefont {Lee}\ \emph {et~al.}(2002{\natexlab{a}})\citenamefont {Lee}, \citenamefont {Jeong}, \citenamefont {Yoon}, \citenamefont {Kim}, \citenamefont {Park},\ and\ \citenamefont {Lee}}]{Lee02}%
  \BibitemOpen
  \bibfield  {author} {\bibinfo {author} {\bibfnamefont {J.}~\bibnamefont {Lee}}, \bibinfo {author} {\bibfnamefont {H.}~\bibnamefont {Jeong}}, \bibinfo {author} {\bibfnamefont {C.}~\bibnamefont {Yoon}}, \bibinfo {author} {\bibfnamefont {C.}~\bibnamefont {Kim}}, \bibinfo {author} {\bibfnamefont {B.~G.}\ \bibnamefont {Park}}, \ and\ \bibinfo {author} {\bibfnamefont {T.~D.}\ \bibnamefont {Lee}},\ }\bibfield  {title} {\enquote {\bibinfo {title} {Interdiffusion in antiferromagnetic/ferromagnetic exchange coupled nife/irmn/cofe multilayer},}\ }\href@noop {} {\bibfield  {journal} {\bibinfo  {journal} {Journal of applied physics}\ }\textbf {\bibinfo {volume} {91}},\ \bibinfo {pages} {1431--1435} (\bibinfo {year} {2002}{\natexlab{a}})}\BibitemShut {NoStop}%
\bibitem [{\citenamefont {Lee}\ \emph {et~al.}(2002{\natexlab{b}})\citenamefont {Lee}, \citenamefont {Kim}, \citenamefont {Yoon}, \citenamefont {Kim}, \citenamefont {Park},\ and\ \citenamefont {Lee}}]{Lee02(2)}%
  \BibitemOpen
  \bibfield  {author} {\bibinfo {author} {\bibfnamefont {J.}~\bibnamefont {Lee}}, \bibinfo {author} {\bibfnamefont {S.}~\bibnamefont {Kim}}, \bibinfo {author} {\bibfnamefont {C.}~\bibnamefont {Yoon}}, \bibinfo {author} {\bibfnamefont {C.}~\bibnamefont {Kim}}, \bibinfo {author} {\bibfnamefont {B.~G.}\ \bibnamefont {Park}}, \ and\ \bibinfo {author} {\bibfnamefont {T.~D.}\ \bibnamefont {Lee}},\ }\bibfield  {title} {\enquote {\bibinfo {title} {Thermal stability of the exchanged biased cofe/irmn electrode for the magnetic tunnel junction as a function of cofe thickness},}\ }\href@noop {} {\bibfield  {journal} {\bibinfo  {journal} {Journal of applied physics}\ }\textbf {\bibinfo {volume} {92}},\ \bibinfo {pages} {6241--6244} (\bibinfo {year} {2002}{\natexlab{b}})}\BibitemShut {NoStop}%
\bibitem [{\citenamefont {Quarterman}\ \emph {et~al.}(2019)\citenamefont {Quarterman}, \citenamefont {Hallsteinsen}, \citenamefont {Dunz}, \citenamefont {Meinert}, \citenamefont {Arenholz}, \citenamefont {Borchers},\ and\ \citenamefont {Grutter}}]{Quarterman19}%
  \BibitemOpen
  \bibfield  {author} {\bibinfo {author} {\bibfnamefont {P.}~\bibnamefont {Quarterman}}, \bibinfo {author} {\bibfnamefont {I.}~\bibnamefont {Hallsteinsen}}, \bibinfo {author} {\bibfnamefont {M.}~\bibnamefont {Dunz}}, \bibinfo {author} {\bibfnamefont {M.}~\bibnamefont {Meinert}}, \bibinfo {author} {\bibfnamefont {E.}~\bibnamefont {Arenholz}}, \bibinfo {author} {\bibfnamefont {J.~A.}\ \bibnamefont {Borchers}}, \ and\ \bibinfo {author} {\bibfnamefont {A.~J.}\ \bibnamefont {Grutter}},\ }\bibfield  {title} {\enquote {\bibinfo {title} {Effects of field annealing on mnn/cofeb exchange bias systems},}\ }\href {\doibase 10.1103/PhysRevMaterials.3.064413} {\bibfield  {journal} {\bibinfo  {journal} {Phys. Rev. Mater.}\ }\textbf {\bibinfo {volume} {3}},\ \bibinfo {pages} {064413} (\bibinfo {year} {2019})}\BibitemShut {NoStop}%
\bibitem [{\citenamefont {Khaydukov}\ \emph {et~al.}(2024)\citenamefont {Khaydukov}, \citenamefont {Dobrynin}, \citenamefont {Hassan}, \citenamefont {Ormston}, \citenamefont {Nikolaev}, \citenamefont {Bencok}, \citenamefont {Fan}, \citenamefont {Steadman}, \citenamefont {Csik},\ and\ \citenamefont {Vorobiev}}]{Khaydukov24}%
  \BibitemOpen
  \bibfield  {author} {\bibinfo {author} {\bibfnamefont {Y.}~\bibnamefont {Khaydukov}}, \bibinfo {author} {\bibfnamefont {A.}~\bibnamefont {Dobrynin}}, \bibinfo {author} {\bibfnamefont {S.}~\bibnamefont {Hassan}}, \bibinfo {author} {\bibfnamefont {M.}~\bibnamefont {Ormston}}, \bibinfo {author} {\bibfnamefont {K.}~\bibnamefont {Nikolaev}}, \bibinfo {author} {\bibfnamefont {P.}~\bibnamefont {Bencok}}, \bibinfo {author} {\bibfnamefont {R.}~\bibnamefont {Fan}}, \bibinfo {author} {\bibfnamefont {P.}~\bibnamefont {Steadman}}, \bibinfo {author} {\bibfnamefont {A.}~\bibnamefont {Csik}}, \ and\ \bibinfo {author} {\bibfnamefont {A.}~\bibnamefont {Vorobiev}},\ }\href {https://arxiv.org/abs/2311.04992} {\enquote {\bibinfo {title} {Depth resolved study of annealing in irmn/(fe, co, cofe) exchange bias systems},}\ } (\bibinfo {year} {2024}),\ \Eprint {http://arxiv.org/abs/2311.04992} {arXiv:2311.04992 [cond-mat.mtrl-sci]} \BibitemShut {NoStop}%
\bibitem [{\citenamefont {Nikolaev}\ \emph {et~al.}(2016)\citenamefont {Nikolaev}, \citenamefont {Pokhil}, \citenamefont {Stankiewicz}, \citenamefont {Patwari},\ and\ \citenamefont {Singleton}}]{Nikolaev16}%
  \BibitemOpen
  \bibfield  {author} {\bibinfo {author} {\bibfnamefont {K.}~\bibnamefont {Nikolaev}}, \bibinfo {author} {\bibfnamefont {T.}~\bibnamefont {Pokhil}}, \bibinfo {author} {\bibfnamefont {A.}~\bibnamefont {Stankiewicz}}, \bibinfo {author} {\bibfnamefont {M.}~\bibnamefont {Patwari}}, \ and\ \bibinfo {author} {\bibfnamefont {E.}~\bibnamefont {Singleton}},\ }\href@noop {} {\enquote {\bibinfo {title} {Sensor structure having increased thermal stability},}\ } (\bibinfo {year} {2016}),\ \bibinfo {note} {uS Patent 9,454,978}\BibitemShut {NoStop}%
\bibitem [{\citenamefont {Han}\ \emph {et~al.}(2023)\citenamefont {Han}, \citenamefont {Song}, \citenamefont {Zhou}, \citenamefont {Ma}, \citenamefont {Ma}, \citenamefont {Gao},\ and\ \citenamefont {Zheng}}]{Han23}%
  \BibitemOpen
  \bibfield  {author} {\bibinfo {author} {\bibfnamefont {Z.}~\bibnamefont {Han}}, \bibinfo {author} {\bibfnamefont {C.}~\bibnamefont {Song}}, \bibinfo {author} {\bibfnamefont {J.}~\bibnamefont {Zhou}}, \bibinfo {author} {\bibfnamefont {Z.}~\bibnamefont {Ma}}, \bibinfo {author} {\bibfnamefont {L.}~\bibnamefont {Ma}}, \bibinfo {author} {\bibfnamefont {H.}~\bibnamefont {Gao}}, \ and\ \bibinfo {author} {\bibfnamefont {F.}~\bibnamefont {Zheng}},\ }\bibfield  {title} {\enquote {\bibinfo {title} {Influence of the deposition conditions on the magnetic properties of fe--co--n thin films},}\ }\href@noop {} {\bibfield  {journal} {\bibinfo  {journal} {Journal of Alloys and Compounds}\ }\textbf {\bibinfo {volume} {934}},\ \bibinfo {pages} {167951} (\bibinfo {year} {2023})}\BibitemShut {NoStop}%
\bibitem [{\citenamefont {Hwang}\ \emph {et~al.}(2016)\citenamefont {Hwang}, \citenamefont {Lee}, \citenamefont {Kim},\ and\ \citenamefont {Kim}}]{Hwang16}%
  \BibitemOpen
  \bibfield  {author} {\bibinfo {author} {\bibfnamefont {T.-J.}\ \bibnamefont {Hwang}}, \bibinfo {author} {\bibfnamefont {J.}~\bibnamefont {Lee}}, \bibinfo {author} {\bibfnamefont {K.~H.}\ \bibnamefont {Kim}}, \ and\ \bibinfo {author} {\bibfnamefont {D.~H.}\ \bibnamefont {Kim}},\ }\bibfield  {title} {\enquote {\bibinfo {title} {Magnetic properties and high frequency characteristics of fecon thin films},}\ }\href@noop {} {\bibfield  {journal} {\bibinfo  {journal} {AIP Advances}\ }\textbf {\bibinfo {volume} {6}} (\bibinfo {year} {2016})}\BibitemShut {NoStop}%
\bibitem [{\citenamefont {Kuo}\ \emph {et~al.}(1998)\citenamefont {Kuo}, \citenamefont {Chang}, \citenamefont {Kuo}, \citenamefont {Yao},\ and\ \citenamefont {Huang}}]{Kuo98}%
  \BibitemOpen
  \bibfield  {author} {\bibinfo {author} {\bibfnamefont {P.}~\bibnamefont {Kuo}}, \bibinfo {author} {\bibfnamefont {S.}~\bibnamefont {Chang}}, \bibinfo {author} {\bibfnamefont {C.}~\bibnamefont {Kuo}}, \bibinfo {author} {\bibfnamefont {Y.}~\bibnamefont {Yao}}, \ and\ \bibinfo {author} {\bibfnamefont {H.}~\bibnamefont {Huang}},\ }\bibfield  {title} {\enquote {\bibinfo {title} {Microstructure and magnetic properties of fecon thin films},}\ }\href@noop {} {\bibfield  {journal} {\bibinfo  {journal} {Journal of applied physics}\ }\textbf {\bibinfo {volume} {83}},\ \bibinfo {pages} {6643--6645} (\bibinfo {year} {1998})}\BibitemShut {NoStop}%
\bibitem [{\citenamefont {Sun}\ and\ \citenamefont {Wang}(2000)}]{Sun00}%
  \BibitemOpen
  \bibfield  {author} {\bibinfo {author} {\bibfnamefont {N.}~\bibnamefont {Sun}}\ and\ \bibinfo {author} {\bibfnamefont {S.}~\bibnamefont {Wang}},\ }\bibfield  {title} {\enquote {\bibinfo {title} {Soft high saturation magnetization fe-co-n thin films for inductive write heads},}\ }in\ \href {\doibase 10.1109/INTMAG.2000.871969} {\emph {\bibinfo {booktitle} {2000 IEEE International Magnetics Conference (INTERMAG)}}}\ (\bibinfo {year} {2000})\ pp.\ \bibinfo {pages} {191--191}\BibitemShut {NoStop}%
\bibitem [{\citenamefont {Wu}\ \emph {et~al.}(2015)\citenamefont {Wu}, \citenamefont {Yang}, \citenamefont {Yang},\ and\ \citenamefont {Zong}}]{Wu15}%
  \BibitemOpen
  \bibfield  {author} {\bibinfo {author} {\bibfnamefont {Y.}~\bibnamefont {Wu}}, \bibinfo {author} {\bibfnamefont {Y.}~\bibnamefont {Yang}}, \bibinfo {author} {\bibfnamefont {Z.}~\bibnamefont {Yang}}, \ and\ \bibinfo {author} {\bibfnamefont {B.}~\bibnamefont {Zong}},\ }\bibfield  {title} {\enquote {\bibinfo {title} {Influence of sputtering power on static and dynamic magnetic properties of fecon films},}\ }\href@noop {} {\bibfield  {journal} {\bibinfo  {journal} {IEEE Magnetics Letters}\ }\textbf {\bibinfo {volume} {6}},\ \bibinfo {pages} {1--4} (\bibinfo {year} {2015})}\BibitemShut {NoStop}%
\bibitem [{\citenamefont {Xu}\ \emph {et~al.}(2011)\citenamefont {Xu}, \citenamefont {Liao}, \citenamefont {Huang}, \citenamefont {Ong},\ and\ \citenamefont {Li}}]{Xu13}%
  \BibitemOpen
  \bibfield  {author} {\bibinfo {author} {\bibfnamefont {F.}~\bibnamefont {Xu}}, \bibinfo {author} {\bibfnamefont {Z.}~\bibnamefont {Liao}}, \bibinfo {author} {\bibfnamefont {Q.}~\bibnamefont {Huang}}, \bibinfo {author} {\bibfnamefont {C.~K.}\ \bibnamefont {Ong}}, \ and\ \bibinfo {author} {\bibfnamefont {S.}~\bibnamefont {Li}},\ }\bibfield  {title} {\enquote {\bibinfo {title} {Influence of sputtering gas pressure on high-frequency soft magnetic properties of fecon thin film},}\ }\href {\doibase 10.1109/TMAG.2011.2151834} {\bibfield  {journal} {\bibinfo  {journal} {IEEE Transactions on Magnetics}\ }\textbf {\bibinfo {volume} {47}},\ \bibinfo {pages} {3921--3923} (\bibinfo {year} {2011})}\BibitemShut {NoStop}%
\bibitem [{\citenamefont {and}, ,\ and\ \citenamefont {and}(2012)}]{Zhang_2012}%
  \BibitemOpen
  \bibfield  {author} {\bibinfo {author} {\bibnamefont {and}}, , \ and\ \bibinfo {author} {\bibnamefont {and}},\ }\bibfield  {title} {\enquote {\bibinfo {title} {Excellent soft magnetic properties realized in fecon thin films},}\ }\href {\doibase 10.1088/1674-1056/21/3/037502} {\bibfield  {journal} {\bibinfo  {journal} {Chinese Physics B}\ }\textbf {\bibinfo {volume} {21}},\ \bibinfo {pages} {037502} (\bibinfo {year} {2012})}\BibitemShut {NoStop}%
\bibitem [{\citenamefont {Meinert}\ \emph {et~al.}(2015)\citenamefont {Meinert}, \citenamefont {B{\"u}ker}, \citenamefont {Graulich},\ and\ \citenamefont {Dunz}}]{Meinert15}%
  \BibitemOpen
  \bibfield  {author} {\bibinfo {author} {\bibfnamefont {M.}~\bibnamefont {Meinert}}, \bibinfo {author} {\bibfnamefont {B.}~\bibnamefont {B{\"u}ker}}, \bibinfo {author} {\bibfnamefont {D.}~\bibnamefont {Graulich}}, \ and\ \bibinfo {author} {\bibfnamefont {M.}~\bibnamefont {Dunz}},\ }\bibfield  {title} {\enquote {\bibinfo {title} {Large exchange bias in polycrystalline mnn/cofe bilayers at room temperature},}\ }\href@noop {} {\bibfield  {journal} {\bibinfo  {journal} {Physical Review B}\ }\textbf {\bibinfo {volume} {92}},\ \bibinfo {pages} {144408} (\bibinfo {year} {2015})}\BibitemShut {NoStop}%
\bibitem [{\citenamefont {Zilske}\ \emph {et~al.}(2017)\citenamefont {Zilske}, \citenamefont {Graulich}, \citenamefont {Dunz},\ and\ \citenamefont {Meinert}}]{Zilske17}%
  \BibitemOpen
  \bibfield  {author} {\bibinfo {author} {\bibfnamefont {P.}~\bibnamefont {Zilske}}, \bibinfo {author} {\bibfnamefont {D.}~\bibnamefont {Graulich}}, \bibinfo {author} {\bibfnamefont {M.}~\bibnamefont {Dunz}}, \ and\ \bibinfo {author} {\bibfnamefont {M.}~\bibnamefont {Meinert}},\ }\bibfield  {title} {\enquote {\bibinfo {title} {{Giant perpendicular exchange bias with antiferromagnetic MnN}},}\ }\href {\doibase 10.1063/1.4983089} {\bibfield  {journal} {\bibinfo  {journal} {Applied Physics Letters}\ }\textbf {\bibinfo {volume} {110}},\ \bibinfo {pages} {192402} (\bibinfo {year} {2017})},\ \Eprint {http://arxiv.org/abs/https://pubs.aip.org/aip/apl/article-pdf/doi/10.1063/1.4983089/14497665/192402\_1\_online.pdf} {https://pubs.aip.org/aip/apl/article-pdf/doi/10.1063/1.4983089/14497665/192402\_1\_online.pdf} \BibitemShut {NoStop}%
\bibitem [{\citenamefont {Dunz}, \citenamefont {Schmalhorst},\ and\ \citenamefont {Meinert}(2017)}]{Dunz17}%
  \BibitemOpen
  \bibfield  {author} {\bibinfo {author} {\bibfnamefont {M.}~\bibnamefont {Dunz}}, \bibinfo {author} {\bibfnamefont {J.}~\bibnamefont {Schmalhorst}}, \ and\ \bibinfo {author} {\bibfnamefont {M.}~\bibnamefont {Meinert}},\ }\bibfield  {title} {\enquote {\bibinfo {title} {{Enhanced exchange bias in MnN/CoFe bilayers after high-temperature annealing}},}\ }\href {\doibase 10.1063/1.5006551} {\bibfield  {journal} {\bibinfo  {journal} {AIP Advances}\ }\textbf {\bibinfo {volume} {8}},\ \bibinfo {pages} {056304} (\bibinfo {year} {2017})},\ \Eprint {http://arxiv.org/abs/https://pubs.aip.org/aip/adv/article-pdf/doi/10.1063/1.5006551/12961278/056304\_1\_online.pdf} {https://pubs.aip.org/aip/adv/article-pdf/doi/10.1063/1.5006551/12961278/056304\_1\_online.pdf} \BibitemShut {NoStop}%
\bibitem [{\citenamefont {Callori}\ \emph {et~al.}(2016)\citenamefont {Callori}, \citenamefont {Rehm}, \citenamefont {Causer}, \citenamefont {Kostylev},\ and\ \citenamefont {Klose}}]{Callori16}%
  \BibitemOpen
  \bibfield  {author} {\bibinfo {author} {\bibfnamefont {S.~J.}\ \bibnamefont {Callori}}, \bibinfo {author} {\bibfnamefont {C.}~\bibnamefont {Rehm}}, \bibinfo {author} {\bibfnamefont {G.~L.}\ \bibnamefont {Causer}}, \bibinfo {author} {\bibfnamefont {M.}~\bibnamefont {Kostylev}}, \ and\ \bibinfo {author} {\bibfnamefont {F.}~\bibnamefont {Klose}},\ }\bibfield  {title} {\enquote {\bibinfo {title} {Hydrogen absorption in metal thin films and heterostructures investigated in situ with neutron and x-ray scattering},}\ }\href@noop {} {\bibfield  {journal} {\bibinfo  {journal} {Metals}\ }\textbf {\bibinfo {volume} {6}},\ \bibinfo {pages} {125} (\bibinfo {year} {2016})}\BibitemShut {NoStop}%
\bibitem [{\citenamefont {Soeya}\ \emph {et~al.}(1994)\citenamefont {Soeya}, \citenamefont {Imagawa}, \citenamefont {Mitsuoka},\ and\ \citenamefont {Narishige}}]{Soeya94}%
  \BibitemOpen
  \bibfield  {author} {\bibinfo {author} {\bibfnamefont {S.}~\bibnamefont {Soeya}}, \bibinfo {author} {\bibfnamefont {T.}~\bibnamefont {Imagawa}}, \bibinfo {author} {\bibfnamefont {K.}~\bibnamefont {Mitsuoka}}, \ and\ \bibinfo {author} {\bibfnamefont {S.}~\bibnamefont {Narishige}},\ }\bibfield  {title} {\enquote {\bibinfo {title} {Distribution of blocking temperature in bilayered ni81fe19/nio films},}\ }\href@noop {} {\bibfield  {journal} {\bibinfo  {journal} {Journal of applied physics}\ }\textbf {\bibinfo {volume} {76}},\ \bibinfo {pages} {5356--5360} (\bibinfo {year} {1994})}\BibitemShut {NoStop}%
\bibitem [{\citenamefont {Devishvili}\ \emph {et~al.}(2013)\citenamefont {Devishvili}, \citenamefont {Zhernenkov}, \citenamefont {Dennison}, \citenamefont {Toperverg}, \citenamefont {Wolff}, \citenamefont {Hj{\"o}rvarsson},\ and\ \citenamefont {Zabel}}]{Devishvili2013}%
  \BibitemOpen
  \bibfield  {author} {\bibinfo {author} {\bibfnamefont {A.}~\bibnamefont {Devishvili}}, \bibinfo {author} {\bibfnamefont {K.}~\bibnamefont {Zhernenkov}}, \bibinfo {author} {\bibfnamefont {A.~J.}\ \bibnamefont {Dennison}}, \bibinfo {author} {\bibfnamefont {B.}~\bibnamefont {Toperverg}}, \bibinfo {author} {\bibfnamefont {M.}~\bibnamefont {Wolff}}, \bibinfo {author} {\bibfnamefont {B.}~\bibnamefont {Hj{\"o}rvarsson}}, \ and\ \bibinfo {author} {\bibfnamefont {H.}~\bibnamefont {Zabel}},\ }\bibfield  {title} {\enquote {\bibinfo {title} {Superadam: Upgraded polarized neutron reflectometer at the institut laue-langevin},}\ }\href@noop {} {\bibfield  {journal} {\bibinfo  {journal} {Review of Scientific Instruments}\ }\textbf {\bibinfo {volume} {84}} (\bibinfo {year} {2013})}\BibitemShut {NoStop}%
\bibitem [{\citenamefont {Vorobiev}\ \emph {et~al.}(2015)\citenamefont {Vorobiev}, \citenamefont {Devishvilli}, \citenamefont {Palsson}, \citenamefont {Rundl{\"o}f}, \citenamefont {Johansson}, \citenamefont {Olsson}, \citenamefont {Dennison}, \citenamefont {Wollf}, \citenamefont {Giroud}, \citenamefont {Aguettaz} \emph {et~al.}}]{Vorobiev2015}%
  \BibitemOpen
  \bibfield  {author} {\bibinfo {author} {\bibfnamefont {A.}~\bibnamefont {Vorobiev}}, \bibinfo {author} {\bibfnamefont {A.}~\bibnamefont {Devishvilli}}, \bibinfo {author} {\bibfnamefont {G.}~\bibnamefont {Palsson}}, \bibinfo {author} {\bibfnamefont {H.}~\bibnamefont {Rundl{\"o}f}}, \bibinfo {author} {\bibfnamefont {N.}~\bibnamefont {Johansson}}, \bibinfo {author} {\bibfnamefont {A.}~\bibnamefont {Olsson}}, \bibinfo {author} {\bibfnamefont {A.}~\bibnamefont {Dennison}}, \bibinfo {author} {\bibfnamefont {M.}~\bibnamefont {Wollf}}, \bibinfo {author} {\bibfnamefont {B.}~\bibnamefont {Giroud}}, \bibinfo {author} {\bibfnamefont {O.}~\bibnamefont {Aguettaz}},  \emph {et~al.},\ }\bibfield  {title} {\enquote {\bibinfo {title} {Recent upgrade of the polarized neutron reflectometer super adam},}\ }\href@noop {} {\bibfield  {journal} {\bibinfo  {journal} {Neutron News}\ }\textbf {\bibinfo {volume} {26}},\ \bibinfo {pages} {25--26} (\bibinfo {year} {2015})}\BibitemShut {NoStop}%
\bibitem [{\citenamefont {Guasco}\ \emph {et~al.}(2022)\citenamefont {Guasco}, \citenamefont {Khaydukov}, \citenamefont {P{\"u}tter}, \citenamefont {Silvi}, \citenamefont {Paulin}, \citenamefont {Keller},\ and\ \citenamefont {Keimer}}]{Guasco22}%
  \BibitemOpen
  \bibfield  {author} {\bibinfo {author} {\bibfnamefont {L.}~\bibnamefont {Guasco}}, \bibinfo {author} {\bibfnamefont {Y.~N.}\ \bibnamefont {Khaydukov}}, \bibinfo {author} {\bibfnamefont {S.}~\bibnamefont {P{\"u}tter}}, \bibinfo {author} {\bibfnamefont {L.}~\bibnamefont {Silvi}}, \bibinfo {author} {\bibfnamefont {M.}~\bibnamefont {Paulin}}, \bibinfo {author} {\bibfnamefont {T.}~\bibnamefont {Keller}}, \ and\ \bibinfo {author} {\bibfnamefont {B.}~\bibnamefont {Keimer}},\ }\bibfield  {title} {\enquote {\bibinfo {title} {Resonant neutron reflectometry for hydrogen detection},}\ }\href@noop {} {\bibfield  {journal} {\bibinfo  {journal} {Nature Communications}\ }\textbf {\bibinfo {volume} {13}},\ \bibinfo {pages} {1486} (\bibinfo {year} {2022})}\BibitemShut {NoStop}%
\bibitem [{\citenamefont {Perrichon}\ \emph {et~al.}(2021)\citenamefont {Perrichon}, \citenamefont {Devishvili}, \citenamefont {Komander}, \citenamefont {P\'alsson}, \citenamefont {Vorobiev}, \citenamefont {Lav\'en}, \citenamefont {Karlsson},\ and\ \citenamefont {Wolff}}]{Perrichon22}%
  \BibitemOpen
  \bibfield  {author} {\bibinfo {author} {\bibfnamefont {A.}~\bibnamefont {Perrichon}}, \bibinfo {author} {\bibfnamefont {A.}~\bibnamefont {Devishvili}}, \bibinfo {author} {\bibfnamefont {K.}~\bibnamefont {Komander}}, \bibinfo {author} {\bibfnamefont {G.~K.}\ \bibnamefont {P\'alsson}}, \bibinfo {author} {\bibfnamefont {A.}~\bibnamefont {Vorobiev}}, \bibinfo {author} {\bibfnamefont {R.}~\bibnamefont {Lav\'en}}, \bibinfo {author} {\bibfnamefont {M.}~\bibnamefont {Karlsson}}, \ and\ \bibinfo {author} {\bibfnamefont {M.}~\bibnamefont {Wolff}},\ }\bibfield  {title} {\enquote {\bibinfo {title} {Resonant enhancement of grazing incidence neutron scattering for the characterization of thin films},}\ }\href {\doibase 10.1103/PhysRevB.103.235423} {\bibfield  {journal} {\bibinfo  {journal} {Phys. Rev. B}\ }\textbf {\bibinfo {volume} {103}},\ \bibinfo {pages} {235423} (\bibinfo {year} {2021})}\BibitemShut {NoStop}%
\bibitem [{\citenamefont {Mishra}\ \emph {et~al.}(2010)\citenamefont {Mishra}, \citenamefont {Radu}, \citenamefont {Valencia}, \citenamefont {Schmitz}, \citenamefont {Schierle}, \citenamefont {D\"urr},\ and\ \citenamefont {Eberhardt}}]{Mishra10}%
  \BibitemOpen
  \bibfield  {author} {\bibinfo {author} {\bibfnamefont {S.~K.}\ \bibnamefont {Mishra}}, \bibinfo {author} {\bibfnamefont {F.}~\bibnamefont {Radu}}, \bibinfo {author} {\bibfnamefont {S.}~\bibnamefont {Valencia}}, \bibinfo {author} {\bibfnamefont {D.}~\bibnamefont {Schmitz}}, \bibinfo {author} {\bibfnamefont {E.}~\bibnamefont {Schierle}}, \bibinfo {author} {\bibfnamefont {H.~A.}\ \bibnamefont {D\"urr}}, \ and\ \bibinfo {author} {\bibfnamefont {W.}~\bibnamefont {Eberhardt}},\ }\bibfield  {title} {\enquote {\bibinfo {title} {Dual behavior of antiferromagnetic uncompensated spins in nife/irmn exchange biased bilayers},}\ }\href {\doibase 10.1103/PhysRevB.81.212404} {\bibfield  {journal} {\bibinfo  {journal} {Phys. Rev. B}\ }\textbf {\bibinfo {volume} {81}},\ \bibinfo {pages} {212404} (\bibinfo {year} {2010})}\BibitemShut {NoStop}%
\bibitem [{\citenamefont {Vallejo-Fernandez}, \citenamefont {Fernandez-Outon},\ and\ \citenamefont {O’Grady}(2007)}]{Vallejo07}%
  \BibitemOpen
  \bibfield  {author} {\bibinfo {author} {\bibfnamefont {G.}~\bibnamefont {Vallejo-Fernandez}}, \bibinfo {author} {\bibfnamefont {L.}~\bibnamefont {Fernandez-Outon}}, \ and\ \bibinfo {author} {\bibfnamefont {K.}~\bibnamefont {O’Grady}},\ }\bibfield  {title} {\enquote {\bibinfo {title} {Measurement of the anisotropy constant of antiferromagnets in metallic polycrystalline exchange biased systems},}\ }\href@noop {} {\bibfield  {journal} {\bibinfo  {journal} {Applied Physics Letters}\ }\textbf {\bibinfo {volume} {91}} (\bibinfo {year} {2007})}\BibitemShut {NoStop}%
\end{thebibliography}%
\end{document}